\documentclass[12pt]{article}
\usepackage{amsmath}
\usepackage{amssymb}
\usepackage{epsfig}
\usepackage{latexsym}
\usepackage{graphics}
\usepackage{cite}
\def\gf{G_\mu}
\def\ord#1{{\cal O}(#1)}
\def\nl{\nonumber\\}

\def\bbe{\bar{\beta}}

\def\alf1{ {\alpha\over\pi} }

\begin{document}
 
\begin{titlepage}
 \begin{flushright}
 {\bf CERN-TH/2000-337}\\
 {\bf DESY-00-179}\\
 {\bf UTHEP-00-0101 }\\
\end{flushright}
 
\begin{center}
{\Large Precision Predictions for (Un)Stable $W^+W^-$ Pair Production\\
 At and Beyond LEP2 Energies$^{\dagger}$
}
\end{center}

\vspace{2mm}
\begin{center}
  {\bf   S. Jadach$^{a,b,c}$,}
  {\bf   W. P\l{}aczek$^{d,c}$,}
  {\bf   M. Skrzypek$^{b,c}$,}
  {\bf   B.F.L. Ward$^{e,f,c}$}
  {\em and}
  {\bf   Z. W\c{a}s$^{b,c}$}

\vspace{2mm}
{\em $^a$DESY-Zeuthen, Theory Division, D-15738 Zeuthen, Germany}\\
{\em $^b$Institute of Nuclear Physics,
  ul. Kawiory 26a, 30-055 Cracow, Poland,}\\
{\em $^c$CERN, Theory Division, CH-1211 Geneva 23, Switzerland,}\\
{\em $^d$  Institute of Computer Science, Jagellonian University,\\
        ul. Nawojki 11, 30-072 Cracow, Poland,}\\
{\em $^e$Department of Physics and Astronomy,\\
  The University of Tennessee, Knoxville, Tennessee 37996-1200, USA,}\\
{\em $^f$SLAC, Stanford University, Stanford, California 94309, USA.}
\end{center}

\vspace{2mm}
\begin{center}
{\bf   Abstract}
\end{center}
We present precision calculations of the processes 
$e^+e^-\rightarrow 4$ fermions, in which the double resonant $W^+W^-$
intermediate state occurs. Referring to this latter intermediate
state as the signal process, we show that, by using the
YFS Monte Carlo event generators YFSWW3-1.14 and KORALW~1.42
in an appropriate combination, we achieve a physical precision
on the signal process, as isolated with LEP2 MC Workshop cuts,
below 0.5\%. We stress the full gauge invariance of our calculations
and we compare our results with those of other authors where appropriate.
In particular, sample Monte Carlo data are explicitly illustrated
and compared with the results of the program 
RacoonWW of Denner {\it et al.}
In this way, we show that the total (physical and technical) 
precision tag for the $WW$
signal process cross section is 0.4\% for 200 GeV, for example.
Results are also given for 500 GeV with an eye toward future linear colliders.

\vspace{5mm}
\begin{center}
{\it Submitted to Phys. Lett. B}
\end{center}
\vspace{10mm}
\renewcommand{\baselinestretch}{0.1}
\footnoterule
\noindent
{\footnotesize
\begin{itemize}
\item[${\dagger}$]
Work partly supported by 
the Maria Sk\l{}odowska-Curie Joint Fund II PAA/DOE-97-316, 
the European Commission Fifth Framework contract HPRN-CT-2000-00149, 
and the US Department of Energy Contracts  DE-FG05-91ER40627 and DE-AC03-76ER00515.
\end{itemize}
}
\begin{flushleft}
{\bf CERN-TH/2000-337}\\
{\bf DESY-00-179}\\ 
{\bf UTHEP-00-0101}\\
{\bf November, 2000}\\
\end{flushleft}

\end{titlepage}
 
The award of the 1999 Nobel Prize for physics to G.~'t Hooft and M.~Veltman, 
and the success of the predictions of their formulation~\cite{gthmv} 
of the renormalized non-Abelian quantum
loop corrections for the Standard Model~\cite{gsw} of the electroweak
interactions in confrontation with data of LEP experiments, 
underscores the need to continue to test 
this theory at the quantum loop level in the gauge boson sector itself. 
This emphasizes the importance of the on-going precision studies of the processes
$e^+e^- \to W^+W^- +n(\gamma)\to 4f+n(\gamma)$
at LEP2 energies~\cite{lep2ybk:1996,frits:1998,frits:1999}, 
as well as the importance of the planned future
higher energy studies of such processes in
LC physics programs~\cite{nlc:1995,jlc:1995,tesla:1998,tesla:2000}.
We need to stress that hadron colliders also have considerable
reach into this physics and we hope to come back to their roles elsewhere~\cite{kkcol1}.

In what follows, we present precision predictions for the
event selections (ES) of the LEP2 MC Workshop~\cite{lep2YR:2000},
for the processes $e^+e^- \to W^+W^- +n(\gamma)\to 4f+n(\gamma)$,
based on our new exact ${\cal O}(\alpha)_{prod}$ YFS-exponentiated
LL ${\cal O}(\alpha^2)$ FSR leading-pole approximation (LPA) formulation,
as it is realized in the MC program 
YFSWW3-1.14~\cite{yfsww3:1998,yfsww3:2000},
in combination with all four-fermion processes 
MC event generator KoralW-1.42~\cite{krlw:1999,krlw:1996b}
so that the respective four-fermion background processes are
taken into account in a gauge-invariant way. Indeed, gauge invariance
is a crucial aspect of our work and we stress that we maintain
it throughout our calculations. Here, FSR denotes final-state
radiation and LL denotes leading-log as usual.

Recently, the authors in Refs.~\cite{racnw:1999} have also 
presented MC program results for the processes
$e^+e^- \to W^+W^- +n(\gamma)\to 4f+n(\gamma),~n=0,1$,
in combination with the complete background processes
that feature the exact LPA  ${\cal O}(\alpha)$ correction.
Thus, we will compare our results, where possible, with those in
Refs.~\cite{racnw:1999} in an effort to check the over-all
precision of our work. As we argue below, the two sets of results
should agree at a level below $0.5\%$ on observables such as the
total cross section.

More specifically, in YFSWW3-1.13~\cite{yfsww3:2000}, the leading-pole 
approximation (LPA) is used to develop a fully gauge-invariant YFS-exponentiated
calculation of the signal process $e^+e^-\rightarrow W^+W^-+n(\gamma)
\rightarrow 4f+n(\gamma)$, which features
the exact ${\cal O}(\alpha)$ electroweak correction to the
production process and the ${\cal O}(\alpha^2)$ LL corrections
to the final-state decay processes. The issue is how to combine this
calculation with that of KoralW-1.42 in Ref.~\cite{krlw:1999,krlw:1996b}
for the corresponding complete Born-level cross section with 
YFS-exponentiated initial-state ${\cal O}(\alpha^3)$ LL corrections.
In this connection, we point out that the LPA enjoys some freedom
in its actual realization, just as does the LL approximation
in the precise definition of the big log L, without spoiling
its gauge invariance. This can already be seen from the
book of Eden {\it et al.}~\cite{eolp:1966}, wherein it is
stressed that the analyticity of the S-matrix applies to the
scalar form factors themselves in an invariant Feynman amplitude,
without any reference to the respective external wave functions and
kinematical (spinor) covariants. The classic example illustrated in
Ref.~\cite{eolp:1966} is that of pion--nucleon scattering,
with the amplitude
\begin{equation}
{\cal M}= \bar u(p_2)[ A(s,t) + B(s,t)(\not\!q_1+\not\!q_2)]u(p_1),
\label{eq1}
\end{equation}
where the $p_i$ are the nucleon 4-momenta, the $q_i$ are the
pion 4-momenta, $u(p)$ is the usual Dirac wave function of the
nucleon, and the invariant scalar functions $A(s,t)$ and $B(s,t)$
of the Mandelstam invariants $s=(p_1+q_1)^2,~t=(q_2-q_1)^2$
realize the analytic properties of the S-matrix themselves
in the complex $s$ and $t$ planes. This means that, whenever we have
spinning particles, we may focus on the analogs of $A$ and $B$
in eq.~(\ref{eq1}) in isolating the respective analytic
properties of the corresponding S-matrix elements.
We note that Stuart~\cite{strt:1995} has emphasized this 
point in connection with the production and decay of
$Z$-pairs in $e^+e^-$ annihilation and in connection with
the production and decay of single $W$'s in $e^+e^-$
annihilation. What this means is that, in formulating the Laurent expansion
of the S-matrix about its poles to isolate the dominant leading-pole term
(the LPA is then realized by dropping all but this leading term), 
we may focus on only
$A$ and $B$, or we may insist that in evaluating the residues
of the poles in the S-matrix the wave functions and kinematical covariants
are also evaluated at the pole positions. When we focus only on
the analogs of $A$ and $B$ in formulating the LPA, we shall
refer to the result simply as the LPA$_a$; when we also 
evaluate the wave functions and
kinematical covariants at the pole positions in isolating the
poles in the analogs of $A$ and $B$ for the LPA, we shall 
refer to the respective
result as the LPA$_b$. As Stuart stressed as well, both the LPA$_a$ 
and the LPA$_b$ are fully gauge-invariant.

For the process under discussion, a general representation is~\cite{strt:1995}
\begin{equation}
{\cal M} = \sum_j \ell_j A_j\left(\{q_kq_l\}\right),
\end{equation}   
where $\{ \ell_j \}$ are a complete set of kinematical covariants
which carry the same transformation properties as does ${\cal M}$,
and the Lorentz scalars $\{q_kq_l\}$ are a complete set of Lorentz
scalar invariants for the external 4-momenta of ${\cal M}$.
In the LPA$_a$, we make a Laurent expansion of the $A_j$ and retain
only their leading poles, without touching the $\{ \ell_j \}$;
in the LPA$_b$, we also evaluate the $\ell_j$ at the position
of the respective leading poles. Evidently, in the latter case,
we must make an analytic continuation of the phase-space point
originally associated with the $\{ \ell_j \}$ to a corresponding
such point for the respective pole positions. See Ref.~\cite{yfsww3:1998}
for an illustration of such a continuation in the context of
the YFS-exponentiated exact ${\cal O}(\alpha)$ calculation
for the production process in $e^+e^- \to W^+W^- +n(\gamma)\to 4f+n(\gamma)$,
and Refs.~\cite{frits:1999} for a similar
illustration in the context of the ${\cal O}(\alpha)$
correction to $e^+e^- \to W^+W^-\to 4f $. Having isolated the appropriate
realization of the LPA at the level
of ${\cal M}$, it must still be decided whether to treat
the phase space used to integrate the cross section exactly or
approximately to match what was done for the $\{ \ell_j \}$
in the case of the LPA$_b$. In all of our work, we stress that
we always treat the exact phase space, both in the LPA$_a$ and in the
LPA$_b$.

In the context of YFS exponentiation, we realize the LPA as follows, as
was briefly described already in Ref.~\cite{yfsww3:1998}.
Taking the respective 4-fermion plus $n$-photon process kinematics
to be as given by (here, $d\tau_{n+4}$ is the respective 
phase space differential with the appropriate normalization):
\begin{equation}
  \label{4fprocess}
  \begin{split}
    &e^-(p_1) + e^+(p_2) 
    \to  f_1(r_1)+\bar{f}_2(r_2) + f'_1(r'_1)+\bar{f}'_2(r'_2) 
    + \gamma(k_1),...,\gamma(k_n)\\
   &\sigma_n = {1\over flux}
    \int d\tau_{n+4}(p_1+p_2;r_1,r_2,r'_1,r'_2,k_1,...,k_n)\\
   &\qquad\qquad
    \sum_{Ferm.~Spin}\; \sum_{Phot.~Spin} 
      |{\cal M}^{(n)}_{4f}(p_1,p_2,r_1,r_2,r'_1,r'_2,k_1,...,k_n)|^2,
  \end{split}
\end{equation}
and that of the corresponding $W^+W^-$ production and decay process
to be as given by
\begin{equation}
  \label{WWprod-decay}
  \begin{split}
   &e^-(p_1) + e^+(p_2) \to  W^-(q_1) + W^+(q_2),\quad \\
   &W^-(q_1) \to f_1(r_1)+\bar{f}_2(r_2),\quad 
    W^+(q_2) \to f'_1(r'_1)+\bar{f}'_2(r'_2),\\
   &\sigma_n = {1\over flux}
    \int d\tau_{n+4}(p_1+p_2;r_1,r_2,r'_1,r'_2,k_1,...,k_n)\\
   &\qquad\qquad
    \sum_{Ferm.~Spin}\; \sum_{Phot.~Spin}
      |{\cal M}^{(n)}_{LPA}(p_1,p_2,r_1,r_2,r'_1,r'_2,k_1,...,k_n)|^2
  \end{split}
\end{equation}
in the context of YFS exponentiation~\cite{yfs:1961,yfsww2:1996},
we proceed according to Refs.~\cite{yfs:1961,yfsww2:1996,yfsww3:1998}
\begin{equation}
  \label{eq:lpa}
  \begin{split}
   &{\cal M}^{(n)}_{4f}(p_1,p_2,r_1,r_2,r'_1,r'_2,k_1,...,k_n)\;\;
    {LPA \atop => }
   {\cal M}^{(n)}_{LPA}(p_1,p_2,r_1,r_2,r'_1,r'_2,k_1,...,k_n) \\
   &=\sum_{Phot.~Partitions}\;
   {\cal M}^{(a),\lambda_1 \lambda_2}_{Prod}(p_1,p_2,q_1,q_2,k_1,...,k_a)\\
   &\times {1\over D(q_1)}\; 
       {\cal M}^{(b-a)}_{Dec_1,\lambda_1}(q_1,r_1,r_2,k_{a+1},...,k_b)\;
    \times {1\over D(q_2)}\; 
       {\cal M}^{(n-b)}_{Dec_2,\lambda_2}(q_2,r'_1,r'_2,k_{b+1},...,k_n),\\
   &D(q_i) = q_i^2-M^2,\quad
    M^2 = (M_W^2-i\Gamma_W M_W)(1-\Gamma_W^2/M_W^2 +{\cal O}(\alpha^3)),\\
   &q_1= r_1+r_2+k_{a+1}+...+k_b;\;\; q_2=r'_1+r'_2+k_{b+1}+...+k_n,
  \end{split}
\end{equation}
so that $M^2$ is the pole in the complex $q^2$ plane
when $q$ is the respective $W$ 4-momentum,
and $M_W$ and $\Gamma_W$ are
the {\em on-shell} scheme mass and width, respectively.
The residues in (\ref{eq:lpa}) are all defined at $q_i^2=M^2$ 
with a prescription according to whether we have LPA$_a$ or LPA$_b$, 
so that (\ref{eq:lpa})
is our YFS generalization of
the formula in eq.~(12) in the first paper in Ref.~\cite{frits:1999}:
\begin{equation}
  \label{eq:enigma}
  {\cal M}^{(n)} = \sum_{\lambda_1,\lambda_2}
    \Pi_{\lambda_1,\lambda_2}(M_1,M_2) 
    {\Delta^+_{\lambda_1}(M_1) \over D_1 }\;
    {\Delta^-_{\lambda_2}(M_2) \over D_2 },~n=0,1,
\end{equation}
where $D_i=D(q_i)$ and $M^2_i=M^2$. We stress that, unlike what is true
of the formula in eq.~(12) in the first paper in Ref.~\cite{frits:1999}
and in eq.~(\ref{eq:enigma}) here, in eq.~(\ref{eq:lpa}) $n$ is arbitrary.
The sum over ``photon partitions'' is over all $10^n$ possible 
attachments of $n$ photons to the six external fermion lines
and the two $W^\pm$ lines (one for the $W$ production and
one for the $W$ decay, respectively).
We make the further approximation that $M^2_i=M^2_W$
in the residues in (\ref{eq:lpa}), always maintaining
gauge invariance, as explained.
Equations (3) and (4) in Ref.~\cite{ward:1987} then give us,
in the presence of renormalization-group-improved perturbation theory,
for the representation
\begin{equation}
\label{eq:real-sum}
 {\cal M}^{(n)}_{LPA}(p_1,p_2,r_1,r_2,r'_1,r'_2,k_1,...,k_n)
 = \sum_{j=0}^\infty 
 {\cal M}^{(n)}_j(p_1,p_2,r_1,r_2,r'_1,r'_2,k_1,...,k_n),
\end{equation}
where ${\cal M}^{(n)}_j$ is the $j$-th virtual photon loop contribution
to the residues in ${\cal M}^{(n)}_{LPA}$, the identifications
\begin{equation}
  \label{eq:virt-sum}
  {\cal M}^{(n)}_j(p_1,p_2,r_1,r_2,r'_1,r'_2,k_1,...,k_n)
  = \sum_{r=0}^j   \mathfrak{m}^{(n)}_{j-r} { (\alpha {B'})^r \over r!},
\end{equation}
where $B'$ is now, for the LPA$_b$ case to be definite, the {\bf on-shell}
virtual YFS infrared function,
which reduces to that given in eqs.~(8) and (9) 
in Ref.~\cite{yfsww2:1996} when we restrict our attention
to the production process, and $\alpha$ is indeed $\alpha(0)$ when it
multiplies $B'$ here.
Let us keep this limit of $B'$
in mind, as we focus on the gauge invariance of
YFSWW3-1.11 in Ref.~\cite{yfsww3:1998}, which treats the 
radiation in the production process,
and on that of YFSWW3-1.13 and YFSWW3-1.14,
in which the radiation from the decay processes is also treated. 
In the LPA$_a$ case, the corresponding $B'$ function is {\bf off-shell}.
Let us discuss first the LPA$_b$ case and comment later on how
the corresponding results for the LPA$_a$ case are obtained.

Here, since the $SU(2)_L\times U(1)$ Ward--Takahashi identities require
(see eq.~(47) in Ref.~\cite{beenakker:1996})
\begin{equation}
\label{eq:wtidnt}
  k^\mu M^{\gamma}_\mu = 0,\quad
  k^\mu M^{Z}_\mu      = i\sqrt{\mu_Z}M^{\chi},\quad
  k^\mu M^{W^\pm}_\mu  = \pm\sqrt{\mu_W} M^{\phi^\pm},\quad
\end{equation}
for $\mu_V$ denoting the squared $V$ boson mass (so that, for $V=W$, $\mu_W=M^2$),
we find that $B'$ is $SU(2)_L\times U(1)$-invariant from the equations in~(\ref{eq:wtidnt})
and our result eq.~(8) in Ref.~\cite{yfsww2:1996}.
From eq.~(\ref{eq:virt-sum}) it then follows that the infrared residuals
$\mathfrak{m}^{(n)}_{j-r}$ are also $SU(2)_L\times U(1)$-invariant.
Here, $\chi$ and $\phi^{\pm}$ are the usual unphysical Higgs fields in
our general renormalizable gauges and we use the notation
of Ref.~\cite{beenakker:1996}, so that $ M^{Z}_\mu$ is their
respective amplitude for the emission of a $Z$ of Lorentz
index $\mu$ and 4-momentum $k$, and $M^{\chi}$ is their corresponding
amplitude for the emission of a $\chi$ with the same 4-momentum, etc.%

Introducing eq.~(\ref{eq:virt-sum}) into~(\ref{eq:real-sum}) gives
\begin{equation}
\label{eq:bexp-sum}
 {\cal M}^{(n)}_{LPA_b}(p_1,p_2,r_1,r_2,r'_1,r'_2,k_1,...,k_n)
 = e^{\alpha B'} \sum_{j=0}^\infty 
   \mathfrak{m}^{(n)}_j(p_1,p_2,r_1,r_2,r'_1,r'_2,k_1,...,k_n).
\end{equation}
Equation~(2.13) in Ref.~\cite{yfs:1961} and eq.~(7) in Ref.~\cite{ward:1987}
then give our $n$-photon
differential cross section, for $P= p_1+p_2$, $\vec P=0$, as%
\begin{equation}
  \label{eq:archaic}
    d\sigma^n_{LPA_b} =
    e^{2\Re \alpha B'} {1\over n!}
    \int \prod_{j=1}^n {d^3k \over k^0_j}
    \delta^{(4)}\left(P-R-\sum_j k_j\right) 
    \left| \sum_{n'=0}^\infty \mathfrak{m}_{n'}^{(n)} \right|^2
    {d^3 r_1  \over r^0_1}    {d^3 r_2  \over r^0_2}
    {d^3 r'_1 \over {r'}^0_1} {d^3 r'_2 \over {r'}^0_2},
\end{equation}
where we note that, when we only focus on the
production process in eq.~(\ref{eq:archaic}), $R$ is the 
produced $WW$ intermediate state; $R=r_1+r_2+r'_1+r'_2$.
Using the second theorem of the YFS program (eq.~(2.15) in~\cite{yfs:1961}), 
we get
\begin{equation}
  \label{eq:m-theory}
  \begin{split}
    \left| \sum_{n'=0}^\infty \mathfrak{m}_{n'}^{(n)} \right|^2 &=
    \tilde S(k_1)\cdots\tilde S(k_n)\bar\beta_0
    +\sum_{i=1}^n\tilde S(k_1)\cdots\tilde S(k_{i-1})\tilde S(k_{i+1})\cdots
    \tilde S(k_n)\bar\beta_1(k_i) 
    \\&
    +\cdots+
    \sum_{i=1}^n\tilde S(k_i)
    \bar\beta_{n-1}(k_1,\cdots,k_{i-1},k_{i+1},\cdots,k_n)
    + \bar\beta_n(k_1,\cdots,k_n),
  \end{split}
\end{equation}
where the real emission function $\tilde{S}(k)$ is given 
by $\tilde{S}_{Prod}(k)$, the real emission infrared function
in eq. (8) in Ref.~\cite{yfsww2:1996} for
{\bf on-shell} $W's$, when we only focus on the emission from the 
production process as we did in Refs.~\cite{yfsww3:1998,yfsww3:2000}.
Since, in general,
\begin{equation}
  \label{eq:s-factor}
  \tilde{S}(k) = \tilde{S}_{Prod}+\tilde{S}_{Dec_1}+\tilde{S}_{Dec_2}+
  \tilde{S}_{Int},
\end{equation}
with
\begin{equation}
  \label{eq:sprd-factor}
  \begin{split}
    \tilde{S}_{Prod}(k) &= -{\alpha \over 4\pi^2}
    \Bigg[ \left( {p_1\over k p_1} - {p_2\over k p_2} \right)^2
    +\left( {{\cal A}q_1\over k{\cal A}q_1}-{{\cal A}q_2\over k{\cal A}q_2}\right)^2 \\
    &+  \left( {p_1\over k p_1} - {{\cal A}q_1\over k {\cal A} q_1}\right)^2
    +  \left( {p_2\over k p_2} - {{\cal A}q_2\over k {\cal A} q_2}\right)^2\\
    &-\left( {p_1\over k p_1} - {{\cal A}q_2\over k {\cal A} q_2}\right)^2
    -\left( {p_2\over k p_2} - {{\cal A}q_1\over k {\cal A} q_1}\right)^2
    \Bigg]\Bigg|_{ ({\cal A} q_i)^2=M_W^2},
  \end{split}
\end{equation}
\begin{equation}
  \label{eq:sdec1-factor}
  \begin{split}
    \tilde{S}_{Dec_1}(k) &= -{\alpha \over 4\pi^2}
    \Bigg[ Q_1Q_2\left( {r_1\over k r_1} - {r_2\over k r_2} \right)^2
    - Q_1Q_W \left( {r_1\over k r_1} - {{\cal A}q_1\over k {\cal A} q_1}\right)^2 \\
    &+ Q_2Q_W  \left( {r_2\over k r_2} - {{\cal A}q_2\over k {\cal A} q_2}\right)^2
    \Bigg]\Bigg|_{ ({\cal A} q_1)^2=M_W^2},
  \end{split}
\end{equation}
\begin{equation}
  \label{eq:sdec2-factor}
  \begin{split}
    \tilde{S}_{Dec_2}(k) &= -{\alpha \over 4\pi^2}
    \Bigg[ Q'_1Q'_2\left( {r'_1\over k r'_1} - {r'_2\over k r'_2} \right)^2
    + Q'_1Q_W  \left( {r'_1\over k r'_1} - {{\cal A}q_2\over k {\cal A} q_2}\right)^2\\
    &-  Q'_2Q_W \left( {r'_2\over k r'_2} - {{\cal A}q_2\over k {\cal A} q_2}\right)^2
    \Bigg]\Bigg|_{ ({\cal A} q_2)^2=M_W^2},
  \end{split}
\end{equation}
\begin{equation}
  \label{eq:int-factor}
  \begin{split}
    \tilde{S}_{Int}(k) &= -{\alpha \over 4\pi^2}
    \Big( {p_1\over k p_1} - {p_2\over k p_2}
    +  Q_1{r_1\over k r_1} - Q_2{r_2\over k r_2}\\
    &+   Q'_1{r'_1\over k r'_1} - Q'_2{r'_2\over k r'_2}\Big)^2
    -\tilde{S}_{Prod}-\tilde{S}_{Dec_1}-\tilde{S}_{Dec_2}
  \end{split}
\end{equation}
is really composed of the scalar product of emission currents 
$\{j_b^\mu(k)\}$ with $k_\mu j_b^\mu(k)=0$,
eq.~(\ref{eq:s-factor}) is also $SU(2)_L\times U(1)$-invariant.
Here
\begin{equation}
\label{eq:continuation}
  {\cal A}q_i \equiv 
  \hbox{\rm analytical continuation of }\; q_i\;
  \hbox{\rm to the point}\; q_i^2=M_W^2.
\end{equation}
This analytical continuation, already described in Ref.~\cite{yfsww3:1998},
does not spoil the gauge invariance, as we see from
Eqs.~(\ref{eq:s-factor},\ref{eq:continuation}). It follows that
the hard-photon residuals $\{\bar\beta_n\}$ are also 
$SU(2)_L\times U(1)$-invariant.

Substitutung (\ref{eq:m-theory}) into (\ref{eq:archaic}) we
finally get the $SU(2)_L\times U(1)$-invariant expression,
which is the fundamental formula of our calculation,
\begin{equation}
  \label{eq:archaic2}
  \begin{split}
    &d\sigma_{LPA_b} =
    e^{2\Re \alpha B' +2\alpha \tilde{B}} {1\over (4\pi)^4}
    \int d^4 y e^{iy(p_1+p_2-q_1-q_2)+D}
    \\&\qquad\qquad\times
    \left[\bar{\beta}_0 +\sum_{n=1}^\infty {d^3 k_j\over k^0_j} e^{-iyk_j}
      \bar{\beta}_n(k_1,...,k_n)\right]
    {d^3 r_1  \over \bar r^0_1}    {d^3 r_2  \over r^0_2}
    {d^3 r'_1 \over \bar {r'}^0_1} {d^3 r'_2 \over \bar {r'}^0_2},
  \end{split}
\end{equation}
where we have defined 
\begin{equation}
\label{eq:ryfsfn}
 D=\int{d^3k\over k_0}
 \tilde S\left[e^{-iy\cdot k}-\theta(K_{max}-|\vec k|)\right],\quad
 2\alpha \tilde B =\int{d^3k\over k_0}\theta(K_{max}-|\vec k|)\tilde S(k). 
\end{equation}
This shows that the parameter $K_{max}\ll \sqrt{s}$ is a dummy parameter on which
$d\sigma$ does not depend. In (\ref{eq:archaic2}), when the complete
value of $\tilde S(k)$ is used, then all $W^{\pm}$ 
radiative effects are contained in the
respective $\bar\beta_n$ residuals, in accordance with the YFS
theory in Ref.~\cite{yfs:1961}, as the non-zero widths of the $W$'s
prevent any IR singularities when a $W$ radiates a 
photon in (\ref{WWprod-decay}).
In our work in YFSWW3, as we indicate below, we make the approximation
of dropping all interference effects between the production and decay
stages and between the two decay stages of (\ref{WWprod-decay}). 
This means that we drop the $\tilde S_{Int}(k)$ in
$\tilde S(k)$ in (\ref{eq:s-factor}) and in (\ref{eq:archaic2}), so that the 
YFS theory then determines the corresponding forms of the
YFS functions $\bar\beta_n$, $B'$ and $D$ as also having the respective
interferences dropped. This approximation, which resums a certain class
of large $W$ radiative effects, corresponds to the YFS exponentiation
of the $W$ production and decay radiation in the LPA,
neglecting of all interferences between the production and decay
stages and between the two decay processes.

We will now comment further on our use of (\ref{eq:archaic2}).
Since the residuals on the RHS (right-hand side) of (\ref{eq:lpa}) 
are on-shell amplitudes in the
$SU(2)_L\times U(1)$ theory, 
for both the production and the decay process, 
it follows that they satisfy the renormalization group 
equations~\cite{c-s:1970,swbg:1973} for the 
$SU(2)_L\times U(1)$ theory, as explained in Ref.~\cite{ward:1987}:
\begin{equation}
  \left(\mu\frac{\partial}{\partial\mu}
    +\beta_j(\{g_{iR}\})\frac{\partial}{\partial g_{jR}} 
    -\gamma_{\Theta_j}(\{g_{iR}\})m_{jR}\frac{\partial}{\partial m_{jR}}
    -\gamma_\Gamma(\{g_{iR}\})\right)\Gamma = 0
  \label{eq:rg}
\end{equation}
for $ \Gamma = {\cal M}^{(n)\lambda_1,\lambda_2}_{Prod},{\cal M}^{(n')}_{Dec,\lambda_i}$, 
where $\mu$ is the arbitrary renormalization point,
$\{g_{iR}\}$ are the respective renormalized $SU(2)_L\times U(1)$ couplings,
and $\{m_{iR}\}$ are the corresponding renormalized mass parameters, etc.,
as defined in Ref.~\cite{swbg:1973}.
It follows that any two schemes for computing 
${\cal M}^{(n)\lambda_1,\lambda_2}_{Prod},{\cal M}^{(n')}_{Dec,\lambda_i}$ 
are related by a finite renormalization-group transformation.
Thus, the complex pole scheme (CPS) and the fermion loop scheme (FLS)
\cite{beenrs}
values of ${\cal M}^{(n)\lambda_1,\lambda_2}_{Prod},{\cal M}^{(n')}_{Dec_i,\lambda_i}$
are related by such a finite renormalization-group transformation.
Specifically, if we denote CPS and FLS values of any quantity $A$
by $A({\rm CPS})$ and $A({\rm FLS})$, respectively,
then we have the identity
\begin{equation}
\label{rgid}
Z_{\Gamma(CPS)}^{-1}\Gamma({\rm CPS})=\Gamma_{un}=
Z_{\Gamma({\rm FLS})}^{-1}\Gamma({\rm FLS}),
\end{equation}
where $\Gamma_{un}$ is the respective unrenormalized value of $\Gamma$
and the $Z_{\Gamma({\rm R})},\;{\rm R}={\rm CPS},\,{\rm FLS}$ 
are the respective field renormalization constants.
It therefore follows that we have the finite renormlization group 
transformation
\begin{equation}
\label{rgtrnfm}
\Gamma({\rm CPS})=
  Z_{\Gamma({\rm CPS})}Z_{\Gamma({\rm FLS})}^{-1}\Gamma({\rm FLS})
\end{equation}
connecting the FLS and CPS schemes. Of course, as the usual implementation
of the FLS scheme omits a gauge invariant set of contributions
to the heavy vector boson widths, for example, in checking the the result
in (\ref{rgtrnfm}) we must omit the corresponding contributions
on both sides of the equation for consistency, as is usualy the case
when we compare renormalzation group improved quantities.
Indeed, as we take the normalization points for the two schemes to be
the complex pole position $M^2=\mu_W$, and as we evaluate 
${\cal M}^{(n)\lambda_1,\lambda_2}_{Prod},{\cal M}^{(n')}_{Dec_i,\lambda_i}$
at this normalization point, the only difference in using
the FLS instead of the CPS will be the approximate treatment in the FLS of the
actual values of 
${\cal M}^{(n)\lambda_1,\lambda_2}_{Prod},{\cal M}^{(n')}_{Dec_i,\lambda_i},~\text{and}~ \mu_W$ 
for the pole position, as already shown in
eq.~(8) of the first paper in Ref.~\cite{frits:1999}: 
thus, for example, if we keep only fermion loops in
${\cal O}(\alpha)$, only the lowest-order width appears in 
eq.~(6) of the first paper in Ref.~\cite{frits:1999};
thus if one would express the resulting
$\mu_W=M^2$ in terms of the respective on-shell mass and
width, only the lowest-order part of $\Gamma_W$ would be given properly.
Similarly, as both schemes normalize at $\mu_W$, 
the difference in the residues 
${\cal M}^{(n)\lambda_1,\lambda_2}_{Prod},{\cal M}^{(n')}_{Dec_i,\lambda_i}$
is that in the FLS only the fermion-loop contributions are retained,
whereas the CPS keeps all loops. 
Evidently, we may extend our ${\cal O}(\alpha)$
calculation in the FLS by using the complete value of $\mu_W$
and including all the one-loop corrections and attendant 
${\cal O}(\alpha)$ real corrections
from Ref.~\cite{ew1}, as we did in Ref.~\cite{yfsww3:1998}. 
We conclude that, after adding in the
entire ${\cal O}(\alpha)$ correction from Ref.~\cite{ew1}, our LPA exact 
${\cal O}(\alpha)_{prod}$ YFS-exponentiated 
calculation arrives at the same amplitudes, independent of whether
we started with the FLS or the CPS.

All of the above results extend directly to the calculation when
we use LPA$_a$ amplitudes, as these are also gauge-invariant
by the gauge-invariance of our leading poles in the S-matrix.
Thus, the only change we must make is that the respective
residues must be calculated in the LPA$_a$  rather than in the LPA$_b$;
for example, in eqs. (\ref{eq:sprd-factor}-\ref{eq:int-factor}), 
$q_i$ would be used instead of ${\cal A}q_i$, etc.
We have done this, as we further illustrate in the following.

Let us now comment on the issue of the pure FSR YFS exponentiation for the
decay processes treated in the LPA. We proceed in analogy
to what is done in Ref.~\cite{yfs3:1992} for the MC YFS3 for the
respective FSR. Specifically, for both decay residue amplitudes
${\cal M}^{(n')}_{Dec_i,\lambda_i}$, we may have contributions
to the respective hard-photon residuals $\bar\beta_n$ due to
emission for the final-state decay processes; in these we
follow the procedure, described in Refs.~\cite{yfs3:1992,ceex2:2000}
and already illustrated in Ref.~\cite{koralz4:1994}, for including these
contributions, using the same YFS methods as we used above for radiative
effects to the initial, intermediate and final states.
Here, we shall neglect all interference effects between
the production and decay processes as we explained above; this is analogous
neglecting all interference effects between the initial and
final states in the $\bar\beta_n$ in Refs.~\cite{yfs3:1992,koralz4:1994}.
We note that it is possible to retain these interference effects, as we have
illustrated in the exact ${\cal O}(\alpha)$ YFS-exponentiated
BHWIDE MC in Ref.~\cite{bhwide} for wide-angle Bhabha scattering
and, more recently, using the new CEEX exponentiation theory
in Ref.~\cite{ceex1:1999}, to all orders in $\alpha$ in the 
new KK MC~\cite{ceex2:2000,ceex1:1999}
for the 2 fermion processes from the $\tau$ threshold to $1$ TeV.
In this way we see that the use of eq.~(\ref{eq:archaic2}) to
include exponentiation of the FSR is fully realizable
by Monte Carlo methods we already tested. 
We stress that, for the same reasons as we gave for the exponentiation of the complete
process, these FSR contributions to the $\bbe_n$
are fully gauge-invariant.

In the current version of YFSWW3, version 1.14, we also 
drop the $\tilde S_{Dec_i}$ terms in $\tilde S$ in eq.~(\ref{eq:s-factor})
and the corresponding terms in the functions $\bar\beta_n$, $B'$ and $D$,
and include FSR using the program PHOTOS~\cite{photos:1994}, which gives
us a LL ${\cal O}(\alpha^2)$ realization of the FSR in which
finite $p_T$ effects are represented as they are in the ${\cal O}(\alpha)$
soft-photon limit. This LL implementation of FSR is fully gauge-invariant.
The ratio of BRs is then used to obtain the ${\cal O}(\alpha)$
correction in the normalization associated with the 
${\cal O}(\alpha)$ correction to the decay processes themselves.
Evidently, these ratios of BRs are also gauge-invariant.
As we illustrate below, for the corresponding non-factorizable
corrections we use an efficient approximation in terms of
the so-called screened Coulomb ansatz~\cite{scc}, which has been
shown to be in good agreement with the exact calculations
for singly inclusive distributions~\cite{scc}. 
This ansatz is gauge-invariant.

We also point out that the current version 1.14 differs from version
1.13 in Ref.~\cite{yfsww3:2000} in that it uses a different renormalization
scheme. Specifically, the scheme used in version 1.13 is the 
so-called $G_\mu$ of Ref.~\cite{ew1}, in which the weak-scale coupling
\cite{ew1} $\alpha_{G_F}$ is used for all terms in the virtual correction,
except those that are infrared-singular, which are given the 
coupling $\alpha \equiv \alpha(0)$. 
In the renormalization-group-improved YFS theory, 
as formulated in Ref.~\cite{ward:1987},
all the terms in the amplitude that involve corrections, in which
the emitted photon of 4-momentum $k$ has $k^2\rightarrow 0$,
should have the coupling strength corresponding to $\alpha(0)$ --
not just those that are IR singular. We therefore have introduced
into YFSWW3 this requirement of the renormalization-group-improved 
YFS theory to arrive at version 1.14. 
We refer to this scheme as our scheme $(A)$. 
According to the renormalization-group-improved YFS
theory, it gives a better representation of the higher
orders effects than does the $G_\mu$ scheme of Ref.~\cite{ew1}.
We stress that this scheme $(A)$ is also gauge-invariant.
The main effect of this change in renormalization scheme between 
versions 1.13 and 1.14 is to change the normalization of version 1.14 
by $\sim -0.3\%$ to $ -0.4\%$ with respect to 
that of version 1.13~\cite{lep2YR:2000}.

The generic size of the resulting shift
in the YFSWW3 prediction, which we just quoted,
can be understood by isolating the well-known
soft plus virtual LL ISR correction to the process at hand, that
has in $\ord{\alpha}$ the expression~\cite{ew1}
\begin{equation}
  \delta^{v+s}_{ISR,LL}=\beta\ln k_0 + 
  \frac{\alpha}{\pi}\left(\frac{3}{2}L+\frac{\pi^2}{3}-2\right),
  \label{eq:s+vISRLL}
\end{equation}
where $\beta\equiv \frac{2\alpha}{\pi}(L-1)$,~$L=\ln(s/m_e^2)$, and
$k_0$ is a dummy soft cut-off that cancels out of the cross section as usual. 
In the $\gf$ scheme of Refs.~\cite{ew1}, which is used in
YFSWW3-1.13, only the part $\beta \ln k_0+(\alpha/\pi)(\pi^2/3)$ of
$\delta^{v+s}_{ISR,LL}$ has the coupling $\alpha(0)$ and
the remaining part of $\delta^{\rm v+s}_{ISR,LL}$ has the coupling
$\alpha_{\gf}\cong\alpha(0)/(1-0.0371)$. 
The renormalization-group-improved YFS theory implies, 
however, that $\alpha(0)$ should be used
for all the terms in $\delta^{v+s}_{ISR,LL}$. This is done in
YFSWW3-1.14 and results in the normalization shift
$\left((\alpha(0)-\alpha_{\gf})/\pi\right)(1.5L-2)$,
whichis $\sim -0.33\%$ at 200\,GeV. 
This explains most of the
change in the normalization of YFSWW3-1.14 with respect to YFSWW3-1.13.
Moreover, it does not contradict the expected total precision tag of
either version of YFSWW3 at their respective stages of testing.
We stress that, according to the renormalization-group equation,
version 1.14 is an improvement over version 1.13 in that it better
represents the true effect of the respective higher-order corrections.

For the purposes of cross checking with ourselves and with 
Ref.~\cite{racnw:1999}, we also created a second scheme,
scheme ($B$), for the realization of the renormalization
in YFSWW3-1.14. In this scheme, we put the entire $\ord{\alpha}$
correction from Refs.~\cite{ew1} at the coupling strength
$\alpha=\alpha(0)$. Since the pure NL hard $\ord{\alpha}$
correction is only $\sim -0.006$ at 200\,GeV, scheme $(B)$
differs in the normalization from scheme $(A)$ by 
$\sim ( \alpha(0)/\alpha_{\gf}- 1)( -0.006) \cong 0.0002$,
which is well below the $0.5\%$ precision tag regime of interest
for LEP2. Thus, scheme $(B)$, which is used in Ref.~\cite{racnw:1999},
is a perfectly acceptable scheme for LEP2 applications.
It gives us a useful reference point from which 
to interpret our comparison with the results of 
RacoonWW from Ref.~\cite{racnw:1999}, which we discuss below.

Having presented our gauge-invariant calculation as it is
realized in the MC YFSWW3-1.14, we will now turn
to illustrating it in the context of LEP2 applications.
Specifically, we always have in mind that one will combine
the cross section from YFSWW3 with that from
KoralW-1.42~\cite{krlw:1999,krlw:1996b} MC,
which is capable to calculate the non-resonant
background contribution in a gauge-invariant way.  
We can do this in two ways, which we will now briefly describe
and refer the reader to Refs.~\cite{v116v151} for the more detailed discussion.
In the first way, we start with LPA$_a$ and we denote the corresponding
cross section from YFSWW3-1.14 as $\sigma(Y_a)$.
It is corrected for the missing background contribution by adding to it
a correction $\Delta\sigma(K)$ from KoralW-1.42 to form
\begin{equation}
\sigma_{Y/K}= \sigma(Y_a)+\Delta\sigma(K),
\label{sigya}
\end{equation}
where $\Delta\sigma(K)$ is defined by 
\begin{equation}
\Delta\sigma(K)= \sigma(K_1)-\sigma(K_3).
\end{equation}
Here, the cross section $\sigma(K_1)$ is the complete 4-fermion result
from KoralW-1.42 with all background diagrams and
with the YFS-exponentiated ${\cal O}(\alpha^3)$ LL
ISR and the cross section $\sigma(K_3)$ is the restricted CC03 
Born-level result from KoralW-1.42 -- again with the YFS-exponentiated
${\cal O}(\alpha^3)$ LL ISR. 
The result in eq.~(\ref{sigya}) is then
accurate to ${\cal O}(\frac{\alpha}{\pi}\frac{\Gamma_W}{M_W})$.
Alternatively, one may start with the cross section for
LPA$_{a,b}$ in YFSWW3-1.14, 
which we refer to as $\sigma(Y_{a})$ and $\sigma(Y_{b})$ correspondingly,
and isolate the respective YFS-exponentiated ${\cal O}(\alpha)$
correction, $\Delta\sigma(Y)$, which is missing from the cross
section $\sigma(K_1)$ as 
\begin{equation}
\label{deltay}
\Delta\sigma(Y_j)=\sigma(Y_j)-\sigma(Y_4),
\end{equation}
where $\sigma(Y_4)$ is the corresponding cross section from YFSWW3-1.14,
with the non-leading (NL) non-ISR ${\cal O}(\alpha)$ corrections to $\bar\beta_n$,
$n=0,1$, switched off. Then the cross section
\begin{equation}
\label{sigky}
\sigma_{K/Y} = \sigma(K_1)+\Delta\sigma(Y_j)
\end{equation}
has the accuracy of ${\cal O}(\frac{\alpha}{\pi}\frac{\Gamma_W}{M_W})$.
We have checked that the results $\sigma_{Y/K}$ and $\sigma_{K/Y}$
are numerically equivalent at the 0.1\% level of interest to us here.
In the following we only show results from the former.
For completeness, we also note that we sometimes identify
$\sigma(Y_1)=\sigma(Y_a),~\sigma(Y_2)=\sigma(Y_b),~\sigma(Y_3)=\sigma(K_3)$,
and $\sigma(K_2)$ is to be identified as the cross section from KoralW-1.42
with the restricted on-pole CC03 Born-level matrix element
with YFS-exponentiated ${\cal O}(\alpha^3)$ LL ISR. This latter
cross section is a future option of KoralW~\cite{kkcol1}.
It would allow further combinations of YFSWW3 and KoralW with the
desired ${\cal O}(\frac{\alpha}{\pi}\frac{\Gamma_W}{M_W})$ accuracy.
Such combinations would be of use in cross checks of our work.

We now illustrate our precision predictions using $\sigma_{Y/K}$.
We have checked that the correction $\Delta\sigma(K)$ is small,
$\lesssim 0.1\%$ for CMS energies $\sim 200$\,GeV. This is summarized
in Tables~\ref{tab:YR-xtot-nocuts} and \ref{tab:YR-xtot-withcuts}, in which we compare the size
of the correction $\Delta\sigma(K)$, labeled $\delta_{4f}$, with
the size of the respective NL non-ISR ${\cal O}(\alpha)$ correction for
the 4-lepton, 2-lepton--2-quark, and 4-quark final states,
with and without the cuts of Ref.~\cite{lep2YR:2000} at 200\,GeV.

Thus, in what follows,
we shall ignore $\Delta\sigma(K)$, as our ultimate 
precision tag objective, $<0.5\%$,
does not require that we keep it. 
It will be analyzed in more detail elsewhere~\cite{kkcol1}.
Further, for the cross section $\sigma(Y_a)$ we have already presented, 
for version 1.13, in Figs.~1-8 of Ref.~\cite{yfsww3:2000},
for the $c\bar s \ell\bar\nu_\ell$, $\ell=e^-,\mu^-$
final states, the $W^{+,-}$ angular distributions in the $e^+e^-$ CMS system,
the $W^{+,-}$ mass distributions,
the distributions of the final-state lepton energies
in the LAB frame ($e^+e^-$ CMS frame), and the final-state lepton
angular distributions in the $W^-$ rest frame their
corrections (relative to the Born-level). 
The main effect  on these differential distributions of the improved 
normalization of version 1.14 
is to shift the normalization, as we discussed above.
Thus, we do not repeat their presentation here.
We refer the reader to the 
results in Ref.~\cite{yfsww3:2000} for an investigation 
into the size of the EW=NL and
FSR effects in the cases listed above
insofar as YFSWW3-1.14 is concerned with the understanding that
the shapes of the distributions apply directly to version 1.14,
but that the normalization of the EW correction should be reduced
by $-0.3\%$ to $ -0.4\%$.
In general, we found in Ref.~\cite{yfsww3:2000} that, depending on the
experimental cuts and acceptances, both the FSR and the EW corrections
were important in precision studies of these distributions;
this conclusion still holds for version 1.14, of course.
For example, in the lepton decay angle distribution,
for the BARE acceptance 
(the final charged lepton is not combined with any photons),
where both the FSR and the EW correction modulate the distribution,
whereas for the CALO acceptance of Ref.~\cite{yfsww3:2000}
(all photons within $5^\circ$ of the
final-state charged lepton are combined with it) the FSR effect is almost
nil whereas the EW correction effect remains at the level of $\sim 2.0\%$.
Here, we focus on 
the total CMS photon energy distribution (Fig.~1a),
and the CMS photon angular distribution (Fig.~1b). 
We show these results both for the BARE and CALO acceptances, as defined in the
4 fermion Section of the 
\underline{Proceedings of the LEP2 MC Workshop}~\cite{lep2YR:2000}.

In Fig.~1a, we see that the total photon energy distributions
are different for the BARE and CALO cases but that the NL non-ISR correction
does not affect them strongly. In Fig.~1b, we
see that, for both the BARE and CALO cases, the NL correction does affect the
photon angular distribution away from the beam directions, as we expect.
Note that this is the NL correction implied by the YFS exponentiation
of our exact ${\cal O}(\alpha)_{prod}$ correction.
Finally, in Fig.~2, we show the effect, in the $W$ mass and angular
distributions, of using the screened
Coulomb correction from Ref.~\cite{scc}, as against the
usual Coulomb correction from Ref.~\cite{ucoul}. 
The effect we see
is a 5 MeV shift in the peak position, associated with the difference
between the screened and usual Coulomb corrections;
we see almost no effect, as expected,
associated with this difference on the $e^+e^-$ CMS W angular distribution.
Since we calculate the finite $p_T$ $n(\gamma)$ corrections to these
distributions, these results are new. Indeed, in Ref.~\cite{lep2YR:2000}
it is shown that the results from RacoonWW and YFSWW3~(Best) for the
distribution in Fig.~1a differ by $\sim 20\%$ and, as we have the dominant
${\cal O}(\alpha^2)$ LL corrections to this distribution whereas in Fig.~20 in
Ref.~\cite{lep2YR:2000} the RacoonWW result only has the exact
${\cal O}(\alpha)$ Born result for the hard photon observable,
we expect that most of this discrepancy would be removed if
the dominant ${\cal O}(\alpha^2)$ LL corrections were included in the
RacoonWW results. This has recently been confirmed in
Ref.~\cite{denner01}, where the authors of RacoonWW show that,
when they include the latter corrections in their
predictions for the $\cos\theta_\gamma$ distribution in
Fig.~1a, the discrepancy is reduced to the level of $\lesssim 5\%$.
In summary, from the results in Ref.~\cite{yfsww3:2000} and
those presented here, we see that the FSR and EW corrections
are necessary for a precision study of the distributions
in the $W$-pair production and decay process at LEP2 energies.

We have made a detailed comparison between our
results and those from Ref.~\cite{racnw:1999} based on the
program RacoonWW in the context of the LEP2 MC Workshop~\cite{lep2YR:2000}.
A complete unpublished preliminary description of the respective
results of this comparison has appeared in Ref.~\cite{lep2YR:2000}.
Here, we focus on the normalization comparison of the
two calculations at LEP2 energies.
We show in Table 3 the comparison of the RacoonWW and YFSWW3-1.14
results for the cross sections as indicated,
without cuts at 200\,GeV (we have looked at the lower energies
184 and 189~GeV and the comparisons there are similar, if not better).
In Table 4, we show the analogous comparisons with the LEP2 MC Workshop cuts
as described in Ref.~\cite{lep2YR:2000}.

We see that for all channels considered, the two sets of results
agree to the level of 0.3\%. 
This gives a total precision estimate of
0.4\% for the theoretical uncertainty on the 200\,GeV CMS energy
$WW$ signal cross-section
normalization when allowance is made for further possible
uncertainties in the higher-order radiative corrections
and the implementation of the LPA~\cite{lep2YR:2000}. 
This is a significant improvement
over the originally quoted $\sim 2\%$ for this uncertainty 
when the NL non-ISR ${\cal O}(\alpha)$
corrections are not taken into account~\cite{ww-talk:lp99}.  
An effort to further reduce this 0.4\% is in progress.

Finally, with an eye toward the LC projects, we have made
simulations using YFSWW3-1.14 for a CMS energy of 500 GeV.
We show our results in Table~\ref{ta:totcsYFSRacLC} for the total cross section
without cuts; here we again compare them to 
the corresponding ones from RacoonWW~\cite{racnw:1999}.
The NL corrections are significant in these results.
Precision studies at LC energies must take these effects into account.
As expected, the percentage difference between YFSWW3-1.14 and
RacoonWW remains below 0.5\% at 500 GeV CMS energy and is somewhat larger
than at 200\,GeV CMS energy. 

In summary, we have presented two recipes for combining YFSWW3 and
KoralW-1.42 to arrive at a gauge-invariant calculation
of the $WW$-pair production and decay in which the
YFS-exponentiated exact ${\cal O}(\alpha)_{prod}$ corrections
are taken into account as well as the ${\cal O}(\alpha^2)$ LL FSR 
and YFS-exponentiated ${\cal O}(\alpha^3)$ LL ISR correction
to the background processes. 
We have illustrated our calculation with several sample MC results
and we have compared our results on the cross-section normalization
with those on Refs.~\cite{lep2YR:2000} at 200\,GeV. In this way, new
precision tag of 0.4\% has been established for this normalization, 
which represents a considerable improvement over the original
result~\cite{ww-talk:lp99} of $\sim 2\%$ when NL non-ISR corrections 
are not taken into account.

\vspace{4mm}
\noindent 
{\large\bf  Acknowledgments}

Two of us (S.J. and B.F.L.W.) acknowledge the
kind hospitality of Prof. A. De R\'ujula and the CERN Theory 
Division while this work was being completed. 
Three of us (B.F.L.W., W.P. and S.J.) 
acknowledge the support of Prof.~D.~Schlatter
and of the ALEPH, DELPHI, L3 and OPAL Collaborations
in the final stages of this work. 
S.J. is thankful for the kind support of the DESY Directorate
and Z.W. acknowledges the support of the L3 Group of ETH Zurich 
during the time this work was performed. All of us thank the
members of the LEP2 MC Workshop for valuable interactions and
stimulation during the course of this work. The authors especially thank
Profs. A.~Denner, S.~Dittmaier and F.~Jegerlehner and Drs.
M.~Roth and D.~Wackeroth for useful discussions and interactions.


\newpage

\begin{table*}[!ht]
\centering
\begin{tabular}{||c|c||c|c||c|c||c||}
\hline\hline
\multicolumn{2}{||c||}{\bf No cuts}  & 
\multicolumn{2}{c||}{$\sigma_{WW}$ [fb]} &
\multicolumn{2}{c||}{$\delta_{4f}$ [\%]} &
\raisebox{-1.5ex}[0cm][0cm]{$\delta_{WW}^{NL}$ [\%]} \\
\cline{1-6}
Final state & Program & Born & ISR & Born &  ISR & \\
\hline\hline
& 
YFSWW3 & 
$219.793 \,(16)$ &
$204.198 \,(09)$ &
--- & --- &
$-1.92 \,(4)$ \\
$\nu_{\mu}\mu^+\tau^-\bar{\nu}_{\tau}$ &
KoralW & 
$219.766\,(26)$ &
$204.178\,(21) $ &
$0.041 $ &
$0.044 $ & 
--- \\
\cline{2-4}
&
(Y$-$K)/Y &
$0.01 \,(1)\% $ &
$0.01 \,(1)\% $ &
--- & --- & --- \\
\hline\hline
 & 
YFSWW3 & 
$659.69 \,(5)$ &
$635.81 \,(3)$ &
--- & --- &
$-1.99 \, (4)$ \\
$u\bar{d}\mu^-\bar{\nu}_{\mu}$ &
KoralW & 
$659.59 \,(8)$ &
$635.69 \,(7)$ &
$ 0.073 $ &
$ 0.073 $ & 
--- \\
\cline{2-4}
&
(Y$-$K)/Y &
$ 0.02 \,(1)\% $ &
$ 0.02 \,(1)\% $ &
--- & --- & --- \\
\hline\hline
 & 
YFSWW3 & 
$1978.37 \, (14)$ &
$1978.00 \, (09)$ &
--- & --- &
$-2.06 \,(4)$ \\
$u\bar{d} s\bar{c} $ &
KoralW & 
$1977.89 \, (25) $ &
$1977.64 \, (21) $ &
$ 0.060 $ &
$ 0.061 $ & 
--- \\
\cline{2-4}
&
(Y$-$K)/Y  &
$ 0.02 \,(1)\% $ &
$ 0.02 \,(1)\% $ &
--- & --- & --- \\
\hline\hline
\end{tabular}
\caption{\sf
The total $WW$ cross sections $\sigma_{WW}=\sigma(K_3),\sigma(Y_4)$ at the Born and ISR level, 
the $4f$ corrections             $\delta_{4f}       =\Delta\sigma(K)/\sigma_{Born}(Y)$ and the
${\cal O}(\alpha)$ NL correction $\delta_{WW}^{NL}=\Delta\sigma(Y_a)/\sigma_{Born}(Y)$ 
at $E_{CM} = 200$\,GeV. 
The numbers in parentheses are the statistical
errors corresponding to the last digits of the results. 
All of the results are without cuts.
}
\label{tab:YR-xtot-nocuts}
\end{table*}

\begin{table*}[!ht]
\centering
\begin{tabular}{||c|c||c|c||c|c||c||}
\hline\hline
\multicolumn{2}{||c||}{\bf With cuts}  & 
\multicolumn{2}{c||}{$\sigma_{WW}$ [fb]} &
\multicolumn{2}{c||}{$\delta_{4f}$ [\%]} &
\raisebox{-1.5ex}[0cm][0cm]{$\delta_{WW}^{NL}$ [\%]} \\
\cline{1-6}
Final state & Program & Born & ISR & Born &  ISR & \\
\hline\hline
& 
YFSWW3 & 
$210.938 \,(16)$ &
$196.205 \,(09)$ &
--- & --- &
$-1.93 \,(4)$ \\
$\nu_{\mu}\mu^+\tau^-\bar{\nu}_{\tau}$ &
KoralW & 
$210.911 \,(26)$ &
$196.174 \,(21)$ &
$0.041 $ &
$0.044 $ & 
--- \\
\cline{2-4}
&
(Y$-$K)/Y &
$ 0.01 \,(1)\% $ &
$ 0.02 \,(1)\% $ &
--- & --- & --- \\
\hline\hline
 & 
YFSWW3 & 
$627.22 \,(5)$ &
$605.18 \,(3)$ &
--- & --- &
$-2.00 \, (4)$ \\
$u\bar{d}\mu^-\bar{\nu}_{\mu}$ &
KoralW & 
$627.13 \,(8)$ &
$605.03 \,(7)$ &
$ 0.074 $ &
$ 0.074 $ & 
--- \\
\cline{2-4}
&
(Y$-$K)/Y &
$ 0.01 \,(1)\% $ &
$ 0.02 \,(1)\% $ &
--- & --- & --- \\
\hline\hline
 & 
YFSWW3 & 
$1863.60 \, (15)$ &
$1865.00 \, (09)$ &
--- & --- &
$-2.06 \,(4)$ \\
$u\bar{d} s\bar{c} $ &
KoralW & 
$1863.07 \, (25) $ &
$1864.62 \, (21)  $ &
$ 0.065 $ &
$ 0.064 $ & 
--- \\
\cline{2-4}
&
(Y$-$K)/Y  &
$ 0.03 \,(2)\% $ &
$ 0.02 \,(1)\%$ &
--- & --- & --- \\
\hline\hline
\end{tabular}
\caption{\sf
The total $WW$ cross sections $\sigma_{WW}=\sigma(K_3),\sigma(Y_4)$ at the Born and ISR level, 
the $4f$ corrections             $\delta_{4f}       =\Delta\sigma(K)/\sigma_{Born}(Y)$ and
${\cal O}(\alpha)$ NL correction $\delta_{WW}^{NL}=\Delta\sigma(Y_a)/\sigma_{Born}(Y)$ 
at $E_{CM} = 200$\,GeV. 
The numbers in parentheses are the statistical
errors corresponding to the last digits of the results. 
All of the results are with the {\em bare} cuts of Sect. 4.1 of 
Ref.~\cite{lep2YR:2000}.
}
\label{tab:YR-xtot-withcuts}
\end{table*}

\begin{table}[!ht]
\begin{center}
  \begin{tabular}{|c|c|c|c|}
    \hline
    \multicolumn{2}{|c|}{\bf No cuts}&
    \multicolumn{2}{|c|}{\bf$\sigma_{\mathrm{tot}}[\mathrm{fb}]$}\\
    \hline
    final state & program & Born & best \nl
    \hline\hline
    & YFSWW3 & 219.770(23) & 199.995(62) \nl
    $\nu_\mu\mu^+\tau^-\bar\nu_\tau$
    & RacoonWW & 219.836(40) & 199.551(46) \nl
    \cline{2-4}
    & (Y--R)/Y & $-0.03(2)$\% &  0.22(4)\% \nl
    \hline\hline
    & YFSWW3 & 659.64(07) & 622.71(19) \nl
    $u\bar d\mu^-\bar\nu_\mu$
    & RacoonWW & 659.51(12) & 621.06(14) \nl
    \cline{2-4}
    & (Y--R)/Y & $0.02(2)$\% &  0.27(4)\% \nl
    \hline\hline
    & YFSWW3 & 1978.18(21) & 1937.40(61) \nl
    $u\bar d s\bar c$
    & RacoonWW & 1978.53(36) & 1932.20(44) \nl
    \cline{2-4}
    & (Y--R)/Y & $-0.02(2)$\% &  0.27(4)\% \nl
    \hline
  \end{tabular}
\end{center}
\caption{\sf Total cross-sections for CC03 from RacoonWW and 
  YFSWW3 at $\sqrt{s}=200$\,GeV without cuts. The numbers in
  parentheses are statistical errors corresponding to the last digits.}
\label{ta:totcsYFSRacnocuts}
\end{table}

\begin{table}[!ht]
\begin{center}
  \begin{tabular}{|c|c|c|c|}
    \hline
    \multicolumn{2}{|c|}{\bf With {\em bare} cuts}&
    \multicolumn{2}{|c|}{\bf$\sigma_{\mathrm{tot}}[\mathrm{fb}]$}\\
    \hline
    final state & program & Born & best \nl
    \hline\hline
    & YFSWW3 & 210.918(23) & 192.147(63) \nl
    $\nu_\mu\mu^+\tau^-\bar\nu_\tau$
    & RacoonWW & 211.034(39) & 191.686(46) \nl
    \cline{2-4}
    & (Y--R)/Y & $-0.05(2)$\% &  0.24(4)\% \nl
    \hline\hline
    & YFSWW3 & 627.18(07) & 592.68(19) \nl
    $u\bar d\mu^-\bar\nu_\mu$
    & RacoonWW & 627.22(12) & 590.94(14) \nl
    \cline{2-4}
    & (Y--R)/Y & $-0.01(2)$\% &  0.29(4)\% \nl
    \hline\hline
    & YFSWW3 & 1863.40(21) & 1826.80(62) \nl
    $u\bar d s\bar c$
    & RacoonWW & 1864.28(35) & 1821.16(43) \nl
    \cline{2-4}
    & (Y--R)/Y & $-0.05(2)$\% &  0.31(4)\% \nl
    \hline
  \end{tabular}
\end{center}
\caption{\sf Total cross-sections for CC03 from YFSWW3 and 
  RacoonWW at $\sqrt{s}=200$\,GeV with {\em bare} cuts of Sect. 4.1
in Ref.~\cite{lep2YR:2000} (see the text). The numbers in 
parentheses are statistical errors
  corresponding to the last digits.}
\label{ta:totcsYFSRaccuts}
\end{table}

\begin{table}[!ht]
\begin{center}
  \begin{tabular}{|c|c|c|c|}
    \hline
    \multicolumn{2}{|c|}{\bf No cuts}&
    \multicolumn{2}{|c|}{\bf$\sigma_{\mathrm{tot}}[\mathrm{fb}]$}\\
    \hline
    Final state & Program & Born & Best \nl
    \hline\hline
    & YFSWW3 & 87.087(11) & 89.607(32) \nl
    $\nu_\mu\mu^+\tau^-\bar\nu_\tau$
    & RacoonWW & 87.133(23) & 90.018(27) \nl
    \cline{2-4}
    & (Y--R)/Y & $-0.05(3)$\% &  $-0.46(5)$\% \nl
    \hline\hline
    & YFSWW3 &261.377(34) &  279.086(97)\nl
    $u\bar d\mu^-\bar\nu_\mu$
    & RacoonWW & 261.400(70)& 280.149(86) \nl
    \cline{2-4}
    & (Y--R)/Y & $-0.01(3)$\% &  $-0.38(5)$\% \nl
    \hline\hline
    & YFSWW3 &783.93(11) & 868.14(31) \nl
    $u\bar d s\bar c$
    & RacoonWW &784.20(21)  & 871.66(27) \nl
    \cline{2-4}
    & (Y--R)/Y & $-0.03(3)$\% &  $-0.41(5)$\% \nl
\hline
\end{tabular}
\end{center}
\caption{\sf Total cross-sections for CC03 from RacoonWW and 
  YFSWW3 at $\sqrt{s}=500$\,GeV without cuts. The numbers in
  parentheses are statistical errors corresponding to the last digits.}
\label{ta:totcsYFSRacLC}
\end{table}


\begin{figure}[!ht]
\centering
\setlength{\unitlength}{0.1mm}
\begin{picture}(1600, 870)
\put( 450, 820){\makebox(0,0)[cb]{\bf (a)} }
\put(1230, 820){\makebox(0,0)[cb]{\bf (b)} }
\put(   0, 0){\makebox(0,0)[lb]{\epsfig{file=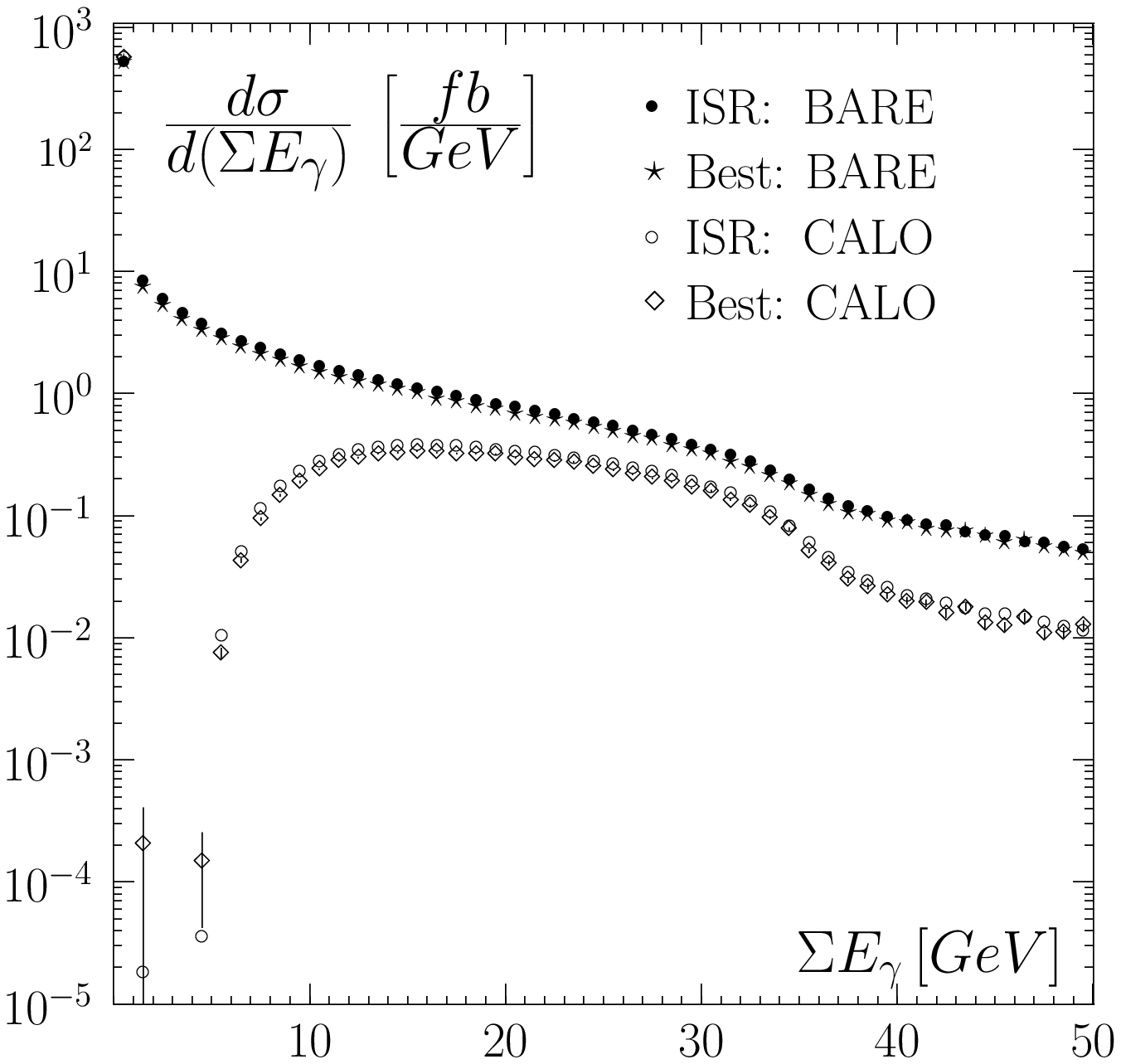,
                                        width=80mm,height=80mm}}}
\put( 800, 0){\makebox(0,0)[lb]{\epsfig{file=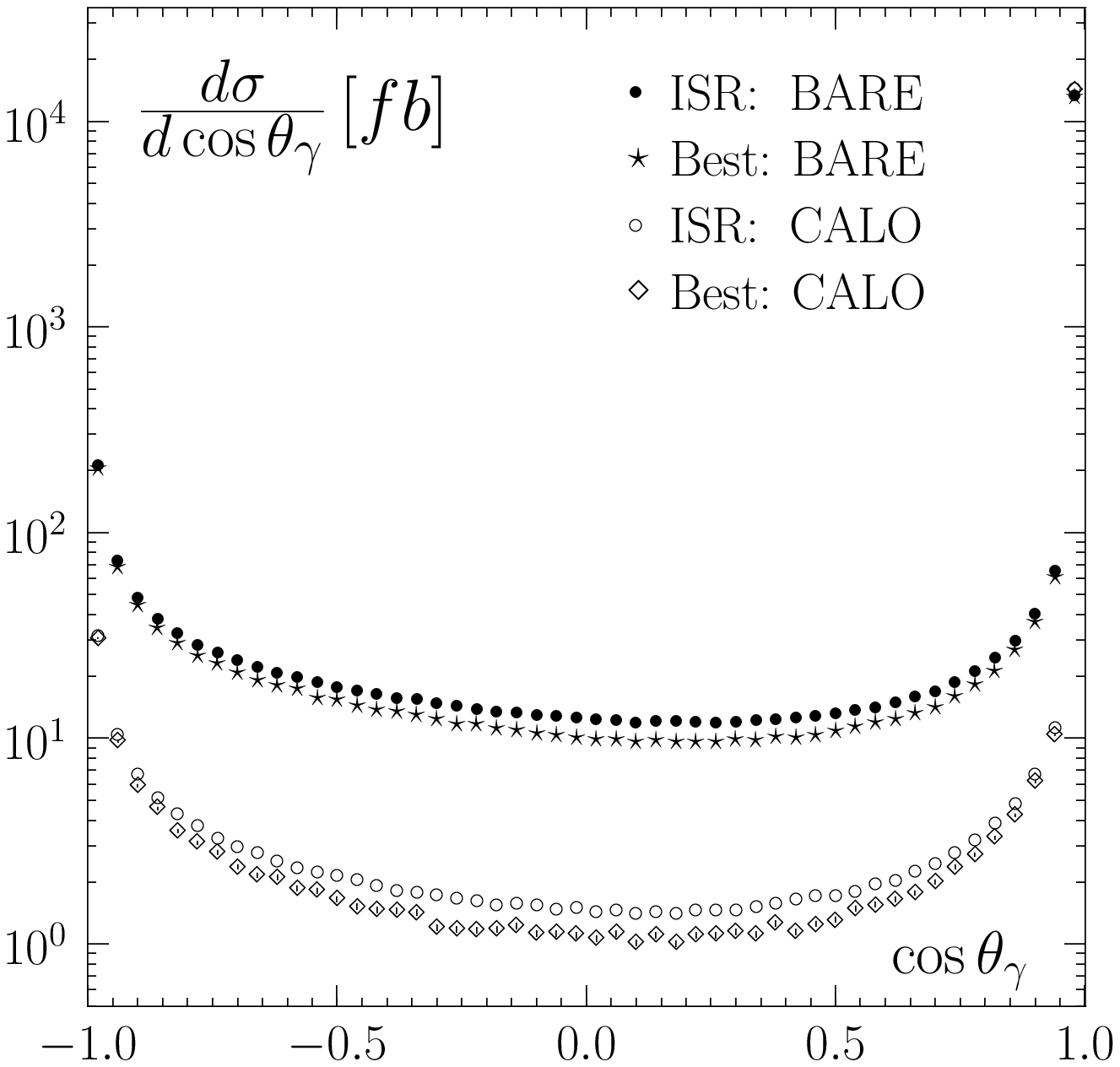,
                                        width=80mm,height=80mm}}}
\end{picture}
\vspace{ -8mm}
\caption{\sf Distributions of the total photon energy (a) and cosine of 
the hardest photon polar angle (b) for the 
$u\bar d+\mu\bar\nu_\mu+n(\gamma)$ final-state. 
The solid, open circle, star,
and diamond curves correspond to the LL BARE, LL CALO,
${\cal O}(\alpha)_{prod}$
${\cal O}(\alpha^2)$ LL FSR BARE and CALO 
YFS-exponentiated results, respectively.
}
\label{fig:1}
\end{figure}

\begin{figure}[!t]
\centering
\setlength{\unitlength}{0.1mm}
\begin{picture}(1600, 870)
\put( 450, 820){\makebox(0,0)[cb]{BARE} }
\put(1250, 820){\makebox(0,0)[cb]{BARE} }
\put(   0, 0){\makebox(0,0)[lb]{\epsfig{file=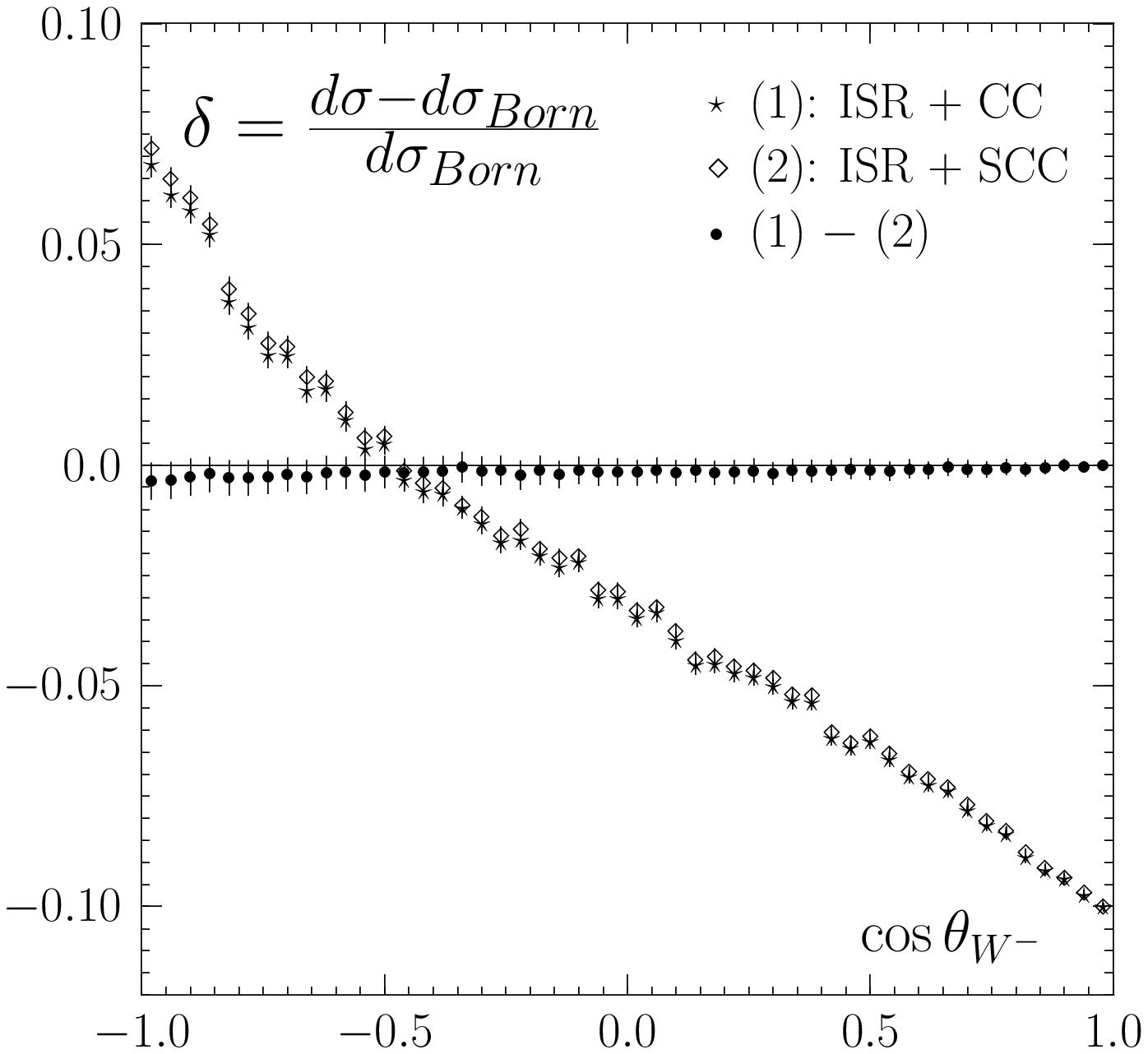 ,width=80mm,height=80mm}}}
\put( 800, 0){\makebox(0,0)[lb]{\epsfig{file=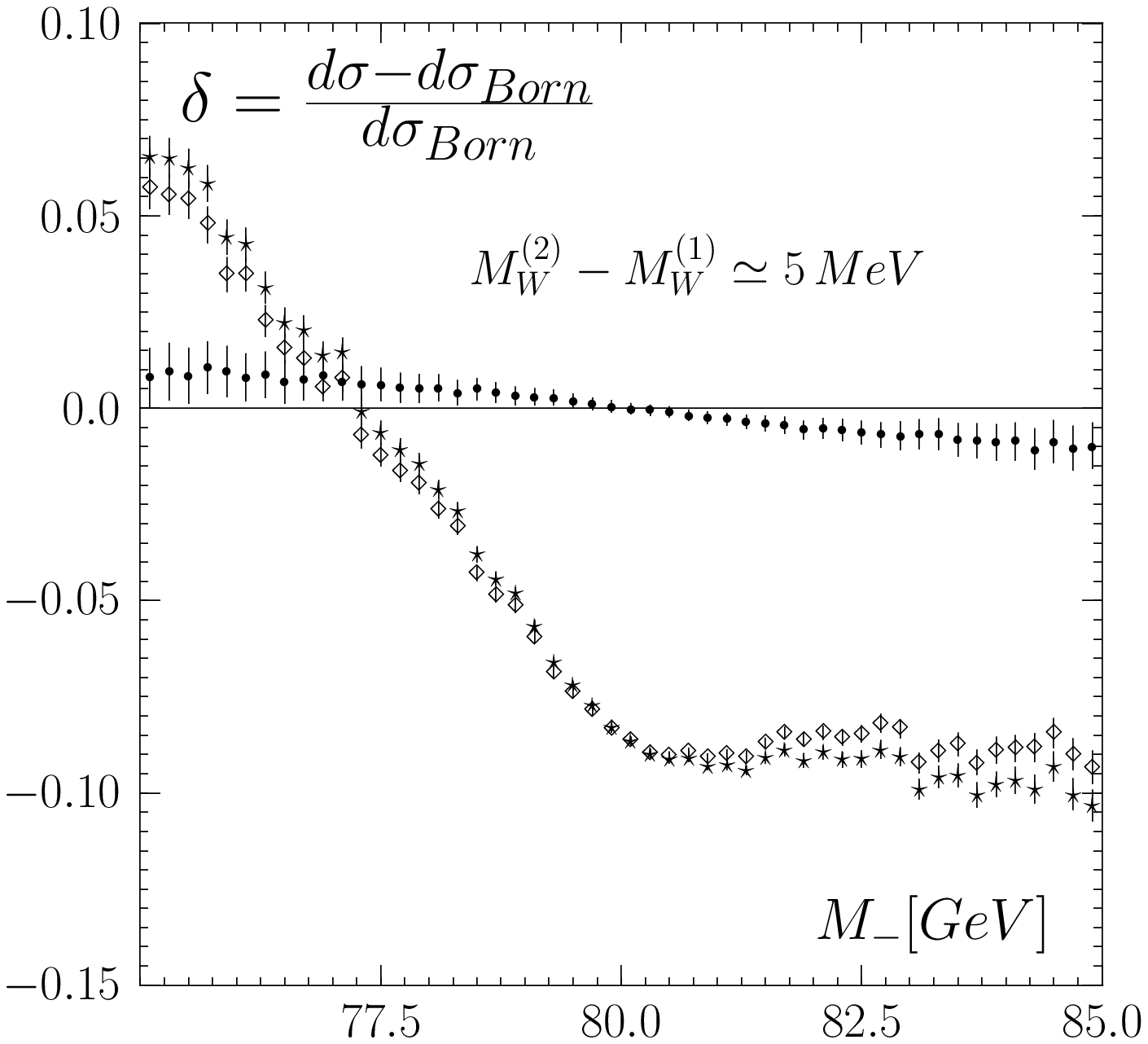     ,width=80mm,height=80mm}}}
\end{picture}
\vspace{ -8mm}
\caption{\sf
  Effects of the screened Coulomb correction (SCC) on the distributions
  of the polar angle (left) and the invariant mass (right) of $W^-$
  in comparison with the usual Coulomb correction (CC) at $E_{CM}=200$\,GeV.
  As indicated the star, solid diamond and large dot curves are the ISR $+$ usual
  Coulomb correction, ISR $+$ screened Coulomb correction and their 
  difference respectively, in the presence of YFS-exponentiation.
  Results are for the 
  $e^+e^- \longrightarrow W^+W^- \longrightarrow u \bar{d} \mu^- \bar{\nu}_{\mu}$ 
  channel. The {\em bare} cut is that of Sect. 4.1 of Ref.~\cite{lep2YR:2000}.
}
\label{fig:2}    
\end{figure}
\vfill

\end{document}